\documentclass[12pt]{article}
\usepackage[margin=1in]{geometry}
\usepackage{algorithm}
\usepackage{algpseudocode}
\usepackage{amsmath}
\title{A Heuristic Method for Scheduling Band Concert Tours}
\author{Linh Nghiem \thanks{Summer Honor Research, University of Miami} 
      \and Advisor: Dr. Tallys Yunes \thanks{Department of Management Science, University of Miami}}
\date{}
\begin{document}
\maketitle
\begin {abstract}
Scheduling band concert tours is an important and challenging task faced by many band management companies and producers. A band has to perform in various cities over a period of time, and the specific route they follow is subject to numerous constraints, such as: venue
availability, travel limits, and required rest periods. A good tour must consider several objectives
regarding the desirability of certain days of the week, as well as travel cost. We developed and
implemented a heuristic algorithm in Java, which was based on simulated annealing, to automatically
generate good tours that both satisfied the above constraints and improved objectives
significantly, when compared to the best manual tour created by the client. Our program also
enabled the client to see and explore trade-offs among objectives while choosing the best tour
that meets the requirements of the business.
\bigskip


\end {abstract}

\hrule
\bigskip
Scheduling, “the allocation of shared resources over time to competing activities,” is an important topic in the Operations Research field that developed from the sequencing of tasks on production machines to the demand of higher efficiency of businesses in general (Yamada \& Nakano 1997; Blazewicz et al. 2014). For example, in the retailing industry, efficient labor scheduling that meets the
fluctuation of customer traffic and sales volume can be a source of competitive advantage in customer service or costs (Quan 2004). In the airline industry, construction of flight schedules, including
but not limited to aircraft routing, manpower planning, and crew scheduling is the key point for all airlines’ optimization on operation (Bazargan 2010). In the sport industry, scheduling promotional events is an “important strategic initiative in attempting to attract fans and corporate sponsors” to the Major League Baseball (MLB) (Fortunao 2006). Today, production scheduling has evolved into an important department in many corporations to ensure the optimum allocation of resources and the efficiency. Many new scheduling methods have also been derived from specific applications
and settings, and have been broadly applied to diverse fields (Blazewicz et al. 2014). 
\par

In this paper, we examined a scheduling problem in a particular context of the music industry. In this industry, concert tours are popular for music bands, singers, and producers to advertise their images to a large number of audience; at the same time, concert tours can be very profitable to band
management companies. Therefore, scheduling concert tour is an important and challenging
task. In fact, the times and locations of performances can have a substantial impact on the cost, revenue, and profit of companies. However, scheduling concert tour usually
requires a tremendous effort, including doing market research about the audience, choosing the cities to
perform, contacting venues to check availability, and so on. Therefore, an algorithm and a computer
program that could put everything together to establish optimal schedules automatically will save much time and increase the efficiency of the business.
\par
The outline of the paper is below: Section 1 describes the problem and compares it with the
Traveling Salesman Problem with Time Windows. Section 2 presents input and output of the
problem. Section 3 models constraints and objectives. Section 4 presents
our methods for solving the problem. Section 5 provides computational results, and Section 6 is
the conclusion of the paper.

\section{Problem Description}
\par Our client was a band management company that created and managed concert tours for the music bands. In a given period, a band had to have one and only one performance in each city. They had to travel from one city to another by bus, and could only travel up to a certain
number of miles per day. Additionally, they could not perform too many performances
consecutively. Moreover, to have a performance in a city on one day, this city had to have at
least one venue available on that day. In addition to satisfying the above constraints, a good tour had to
consider many business objectives: The tour should have as many performances on Thursday and Friday, as few
performances on Monday and Tuesday, and had as few travel miles as possible. We defined Thursday and
Friday as ``good'' days for performance, and Monday and Tuesday as ``bad'' days for performance. For simplicity, we referred to the
number of performances on good days in a tour as \textbf{the number of good days} of this tour, and referred to
the number of performances on bad days in a tour as \textbf{the number of bad days} of this tour. \par
Before giving the problem to us, the client had created tours manually. The client gave us a sample data (described in the next section). The best tour that the client created from this data had a low number of miles, but it had too many bad days and too few
good days. The client asked us if we could generate better tours automatically and see trade-offs among objectives while choosing the best tour.\par
The Traveling Salesman Problem with Time Windows (TSPTW) was related to our problem. The
fundamental idea of TSPTW is to find a tour that visits each node within a given time window that
minimizes total travel distance (Savelsbergh 1985). In our problem, each city had to be visited only
once and on a day that had, at least, one available venue. The time window, therefore, was not a single
continuous time interval, but the union of many countable intervals. Additionally, the tour was subject
to other constraints, including required rest periods and travel limit. It also had to consider two additional
objectives, including the number of good days
and the number of bad days. Therefore, the contribution of this paper is to present a
computational method that combines several simple algorithms to solve a specific extended
problem of the TSPTW.
\section{Modeling Input and Output\protect\footnote{Data and program are available upon request through lnghiem@smu.edu.}}
\subsection{Input}
We received the following data and specific requirement from the client, which resulted from
experience and the nature of the business:
\begin{itemize}
	\item Availability File: This was an Excel file that indicated the availability of each venue in many cities over a
period. Specifically, the file had the availability data of venues in 15 cities over 42 days, from
Monday, Oct-14 to Sunday, Nov-24. On a day, each venue had one of the following statuses:
open (o), meaning absolutely available to reserve; confirmed (c), meaning absolutely
unavailable to reserve; open/hold (o/h), meaning likely availability; pending (p), meaning the
venue was in contract with another artist, so was unavailable; xh (``h'' meant waitlist, x was a number that indicated the
position of the client on the waitlist); or empty, meaning no information, so not available.
	\item Travel limit: The band had to travel no more than 500 miles per day.
	\item Required rest period: The band could have no more than 4 consecutive performances.
\end{itemize}
Then we processed the data as followings:
\begin{itemize}
	\item We numbered each city from $1$ to $m$ 
	and we numbered each date from $0$ to $n-1$ from the earliest to the latest date in the period.
	
	\item We computed a distance matrix among the cities in terms of miles (\textit{Mile} matrix). In Microsoft
Excel, we used a user-created function that used Google Map to calculate distance between two
cities. We had 15 cities, so the Mile matrix had dimension 15 by 15. Because the distance was calculated
based on the real distance by traffic, the matrix was not symmetric. For example, the distance
from Miami to Tampa was not the same as the distance from Tampa to Miami in miles.
We referred to the total travel distance of a tour in miles as the \textbf{total mile distance} of this
tour. We used the notation \textit{Mile[$i$][$j$]} to refer to the distance in miles to travel from city $i$ and city $j$.
	\item We computed a distance matrix among the cities in terms of days to get a \textit{Day} matrix. Each element
of the Day matrix was calculated by dividing the respective element of the \textit{Mile} matrix by
500 (the upper limit number of miles per day). We then rounded it down to the nearest integer number.
Unlike the Mile matrix, the Day matrix was symmetric, because the two-way traffic distances
did not differ more than 500 miles. Each element in the Day matrix represented the desirable
time for the band to travel from one city to another. We referred to the total travel distance of a tour in terms
of days  as \textbf{the total day distance} of this tour. We used the notation \textit{Day[i][j]} to refer to the desirable number of days for the band to travel from city $i$ to city $j$.
    \item We categorized the availability of the venues based on the following rules:
    	\begin {itemize}
    		\item If the status of a venue was either ``o'', ``o/h'', ``1h'', ``2h'', or ``3h'', it indicated availability. The client told us that he
was likely to get the venue if his position in the waitlist was 3 or less, so we
considered ``1h'', ``2h'', and ``3h'' as available.
			\item If the status of a venue was either ``c'' or blank, it indicated absolute unavailability. The client could not get the
venue that had been absolutely unavailable in any case.
			\item The remaining statuses were ``4h'', ``5h'', or more, indicating a high position of the client in the
waitlist. We considered them relative unavailability, meaning that the client was still able
to get the venue if needed. For example, in Los Angeles, the client was likely to
get the venue with ``5h.''
			\end{itemize}
			
	\item Finally, we established \textit{Availability matrix} as followings:
	\begin{itemize}	
	\item Usually, one city had only one venue, so the availability of a city on a day was the same as the availability of the venue there. In case a city had more than one venue (like Boston and New York in the data), we would merge all of its venues
			to indicate the availability of the city, because the band had to perform only once in one city. A city would be
			considered available on one day if at least one venue there was available on that day. A city would be
			considered absolutely unavailable on one day if all venues there were absolutely unavailable on that
			day. Otherwise, a city would be considered relatively unavailable. 
	\item The codes for the availability of the cities were as following: 1 for available, -1 for absolutely unavailable, and 0 for relatively unavailable. 	
	\item We used the notation \textit{Availability[$i$][$j$]} to indicates the status of city $j$ on day $i$.	
			
		\end{itemize}
\end{itemize}

\subsection{Output}
Given the input data, we attempted to produce tours that went through all cities over the period of
time. We represented each tour by a vector whose length was equal to the number of days in the given data. In
other words, our solution was expressed in the form:
\begin{equation}
  v = (v[0], v[1], ... , v[n-1])
\end{equation} 
If $v[i] = 0$, there was no performance on day $i$; the band could travel or had a break in day i. If $v[i] =
k > 0$, the band would perform at city $k$ on day $i$, $ k = 1,2,..., m $, and $ i = 0, 1, ..., n-1$.
\section{Modeling Constraints and Objectives}
\subsection{Absolute approach}
A tour was acceptable in strict sense if all the following constraints were
satisfied:
\begin{itemize}
	\item Availability requirement: For each $i$, if $v[i] = k > 0$ then $Availability[i][k-1] = 1;$ otherwise,
there would be an availability violation at day $i$.
	\item Break Requirement: The band had to have a break after $X$ days: For each $i = 1,2, … n -− X$, the
product $P(i, X) = v[i] v[i + 1]. . . v[i + X] = 0$. If $P(i, X) \neq 0$, there would be a break violation at
day $i$. In our situation, $X = 4$.
	\item Separation Requirement: The band had to take a greater than or equal to the number of desirable days to
travel from one city to another. That meant, for each $i$, if $v[i] \neq 0$, let 
	\begin{equation}
  j = min\{k: k > i, v[k] \neq  0\} 
  	\end{equation}
Then
	\begin{equation}  	
	j -− i \geq Day [v[i] −- 1][v[j] −- 1]]
    \end{equation}
Otherwise, there would be a separation violation at day $i$.
\end{itemize}
Associated with the above constraints, a strict cost function  used to compare the tours had the following form:
	\begin{equation}
F(v) = a_1x_1 + a_2x_2 + a_3x_3 = \sum_{i=1}^{3} a_ix_i
    \end{equation}
where \\
$a_i$ = weight of objective $i$\\
$x_1$ = total mile distance of the tour \\
$x_2$ = total number of good days of the tour\\
$x_3$ = total number of bad days of the tour

\noindent We aimed to find a tour that minimized the cost function, so while $a_1$ and $a_3$ were positive, $a_2$ was
negative. We might choose different values for $(a_1, a_2, a_3)$ based on the relative importance of each
objective. \par

The problem with the absolute approach was that it worked on a very limited number of data. Besides
the data the client gave us, we also simulated some random Availability matrices to ensure our algorithm
could work for all data. Our method below contained the construction of an initial solution; however,
no initial solution was found in most cases when we used the absolute approach.
\subsection{Relaxed Approach}
The relaxed approach considered any tour acceptable, but penalized violation of constraints in the
cost function. However, regarding the availability and separation requirement, all violations were
not penalized equally. We divided availability violations into 2 types: A type 1 availability violation would be
counted if a performance was placed on a venue that was relatively unavailable, while a type 2
availability violation would be counted if a performance was placed on a venue that was absolutely unavailable.
A type 2 availability violation was more penalized than a type 1 availability violation. \par
Similarly, not all separation violations should be treated equally. We also categorized separation violations
into 2 types: A type 1 separation violation would be counted if the band had to take exactly 1 day fewer than
the desirable number of days to travel between two cities, whereas a type 2 separation violation would be counted if the
band had to take more than 1 day fewer than the desirable time to travel. For example, the desirable
number of days to travel from Los Angeles to Atlanta was 7 days; if the band had to travel in 6 days, it would be a
type 1 separation violation. If the band had to do it in 4 days, it would be a type 2 separation violation.
A type 2 separation violation was also more penalized than a type 1 separation violation. \par

Our relaxed cost function had the following form:
\begin{equation}
F(v) = \sum_{i=1}^{3} a_ix_i + \sum_{j=1}^{5}b_jx_j 
\end{equation}
where \\
$a_i$ = weight of objective $i$ \\
$x_1$ = total miles of the tour \\
$b_j$: weight of penalties for constraint violations $j$ \\
$y_1$: the number of type 1 availability violations \\
$y_2$: the number of type 2 availability violations \\
$y_3$: the number of break violations \\
$y_4$: the number of type 1 separation violations\\
$y_5$: the number of type 2 separation violations \par
\noindent We aimed to find a tour that minimized the relaxed cost function. Similar to the absolute approach,
we could choose different values of $a_i$ and $b_i$ based on the relative importance of each objective and
violation.
\section{Method}
Given difficulties in finding solutions for the absolute approach, we used the relaxed approach that
imposed no constraints in constructing the initial tour and attempted to minimize the relaxed cost function. We
explored heuristic methods to generate different tours in a reasonable amount of time. We began
with some heuristic algorithms to produce an initial tour and then improved it by simulated
annealing.
\subsection{Construction Phase}
Although all tours were acceptable in the relaxed approach, we still wanted to begin at a relatively
good tour, meaning a tour did not have too many violations. In the construction phase, we first
used the Nearest Neighbor algorithm to create a list of city order that minimized total day distance,
then put the list in the tour to satisfy separation requirement, and finally did some backward swaps
to reduce as many availability violations as possible.
\paragraph{Initial Construction Using Nearest Neighbor}
The Nearest Neighbor (NN) algorithm is one of the most fundamental algorithms in operations
research to solve the classic Traveling Salesman problem, which is to find the cheapest way of
visiting all the cities given a set of cities and cost of traveling between each pair of them (Applegate
et al. 2011). Note that the cost of traveling may include distance, time, and other costs. For the
Traveling Salesman Problem, the NN algorithm is usually criticized for the computational inefficiency
in case of big data, because it requires a large amount of memory to store all information (Malkov et
al. 2014). However, because the goal of using the NN in our case was just an intermediate step to
create a tour to begin (we did not evaluate how good it was), we still apply the NN algorithm to
the list of cities where the band had to perform.

Applying the general NN algorithm as described by Shakhnarovich et al. (2006), we implemented the NN algorithm for
this problem was expressed briefly in the following steps:
\begin{itemize}
\item Step 1: Start at a random city, set it as the current city.
\item Step 2: Find the next city C that is unvisited and closest to the current city in terms of days. If
two cities has the same nearest distance from the current city, pick the city coded with a smaller
number.
\item Step 3: Set current city to C, and mark C is visited.
\item Step 4: Repeat step 2 and 3 until all cities are visited
\end{itemize}
Because we started a random city, the list after Step 4 would not have the least total day distance. We
tried two-exchange moves that swapped two cities in the list to see if the total day distance could be
further reduced.
\begin{itemize}
\item Step 5: Choose a pair of cities in the list, and swap two chosen cities.
\item Step 6: After swapping, recalculate the total travel distance of the list.
\item Step 7: If the total travel distance is reduced, keep the list; otherwise, undo the swap.
\item Step 8: Repeat step 5 to 7 until all pairs are chosen.
\end{itemize}
After step 8, we had a list of city order with the least total mile distance. 
\paragraph{Separation Requirement}
We started building a tour from the above list by dealing with the separation requirement in the following steps:
\begin{itemize}
\item Step 9: Put the first city in the list order on the first day of the tour
\item Step 10: Set the current day CD be 0 and the current city CC be the first city in the list.
\item Step 11: Put the next city NC in the city order in a position $j$ that the difference between $j$ and CC
equals the desirable time to travel between NC and CC.
\item Step 12: Update CD to $j$, and update CC to NC
\item Step 13: Repeat Step 11 and 12 until all cities in the city order are put in the tour
\item Step 14: If no city is put on one day, set the respective position in the tour be 0.
\end{itemize}
\paragraph{Reduction of Availability Violation through Backward Swap}
After step 14, because the total day distance was usually less than the total number of days of the period,
all performances were condensed in the early days. For example, the following tour had all
performances in the first 21 days (out of 42 days):

\noindent[14, 12, 13, 15, 0, 11, 10, 0, 3, 2, 1, 0, 4, 5, 6, 0, 0, 7, 8, 0,
9, 0, 0, 0, 0, 0, 0, 0, 0, 0, 0, 0, 0, 0, 0, 0, 0, 0, 0, 0, 0, 0] 

However, we have yet imposed any availability requirement in constructing the above tour, so the tour
was likely to have many availability violations. In fact, the above tour had 7 availability violations, so we attempted to reduce it. We checked each city and see if it was available on the day that the tour visited. If
it was not, we would attempt to switch it with one available day that currently be 0 at the end of the
tour (after the current last performance), because the further a day was, the more likely that venues were available on
that day. For example, city 14 was not available on the day 0 (the first day), but available on the day 41
(the last day), which was currently 0. We, therefore, swapped city 14 from day 0 to day 41. After
doing the backward swap, we got the following tour with only 5 availability violations:

\noindent [0, 12, 13, 15, 0, 11, 10, 0, 3, 2, 1, 0, 4, 5, 6, 0, 0, 7, 8, 0, 0,
0, 0, 0, 0, 0, 0, 0, 0, 0, 0, 0, 0, 0, 0, 0, 0, 9, 0, 0, 0, 14] 

The cities now were more likely to spread out over the tour. More importantly, however, we got the
tour that had fewer availability violations.
\subsection{Improvement Phase through Simulated Annealing} 
We called the tour after doing backward swap as the initial solution for the improvement phase. In constructing the initial solution, we did not pay attention to the objectives (good/bad days, total mile distance); we only attempted to get a tour with as few violations as possible. In the improvement phase, we
improved the initial solution based on the cost, and chose the tour that minimized the
cost in our allowed time.
\paragraph{Local Search}
We started at an initial solution, and then for each current solution, we iteratively searched its
neighborhood in the solution space. In the pure local search algorithm, we would go to a new solution
only if its cost was less than the cost of the current solution. In this problem, we defined the neighbor
of a solution S as a set of all tours that could be generated through a \textit{feasible two-exchange move}, one that swapped
either two random performances or one performance with one rest day in the solution S.
\paragraph{Simulated Annealing}
The major disadvantage of the pure local search algorithm was that it was likely to be trapped in the local
optimum, which might not be the global optimum, because it only moved when there was a
decrease in cost. Simulated Annealing, developed by Metropolis et al. [1953], resolved the disadvantage by
allowing a move with an increase in cost to get out of the local minimums with some probability that was
controlled by an acceptance probability function. Because of its simplicity and ease of implementation, simulated annealing is a powerful and popular heuristic method in a wide variety of
scientific disciplines.

The pseudo-code of the simulated annealing algorithm (Trick, Yildiz, \& Yunes 2012) is given below. For the meaning of each parameter in the simulated annealing, please refer to the Chapter 2 of
\textit{Modern Heuristic Techniques for Combinatorial Problems} by Reeves (1993).

\begin{algorithm}
\caption{Simulated Annealing}
\label{SA}
\begin{algorithmic}
	\While{time limit not exceeded}
		\State $S$ = initial tour;
		\State bestTour = S;
		\State $t=t_0$;
	 	\While {$t > TEMPLIMIT$}
				\For{all ITER iterations}
						\State Pick a random feasible two-exchange move E
						\State $d$ = Impact of E in cost function
						\If{$d < 0$}
						   \State Execute E
						   \If{cost of new solution is less than cost of bestTour}
							  \State Update bestTour 
					        \EndIf
						\Else
							   \State $x$ = random number in [0,1]
							   \If{$x < exp(-d/t)$}
							   		  \State Execute E
							   \EndIf
						\EndIf	   		 	  	
				\EndFor
				\State $t=t*ALPHA$
	\EndWhile
  \EndWhile	
\end{algorithmic}  
\end{algorithm}

\section{Parameter Experiment and Computational Results}
We experimented different sets of weights and penalties for the relaxed cost function to get the
tours with different emphases in objectives and constraints. We coded the program in Java, and the
program was run in a Window PC with Intel Core i5 2.4 GHz processor. The length for each run was
30 minutes.

The relaxed cost function was used: 
\begin{equation*}
F(v) = \sum_{i=1}^{3} a_ix_i + \sum_{j=1}^{5}b_jx_j 
\end{equation*}

\par We started with $a_1 = -200$, $a_2 = 200$, $a_3 = 20,$ meaning that we were willing to increase the
total miles by 10 miles in exchange for either one more good day or one fewer bad day. Moreover, any type 2 violation should be much more
penalized than any type 1 violation. We set $b_1 = 10,000$, $b_2 = 1,000,000$, $b_3 = 10,000$, $b_4 = 10,000$, and
$b_5 = 2,000,000$.
We used the following values for the parameters of simulated annealing:
\bigskip

\begin{tabular}{l c}
\hline
Parameter & Value \\
\hline
Initial temperature $t_0$ & 5,000\\
Temperature limit TEMPLIMIT & 500\\
Number of iterations ITER & 5000\\
ALPHA $\alpha$ & 0.95 \\
\hline
\end{tabular}

\bigskip
After running the program for 30 minutes, all tours we produced, including the immediate and the
best ones, had all zero violations. This demonstrated that our choice of penalties were good. The
best tour had 6 good days, 3 bad days, and 7,960 miles. Although the manual tour that the client had created had
the same total miles, it had only 4 good days and 6 bad days. Our program, therefore, enhanced the
objective regarding the desirability of performance day significantly.

For the second run, we kept all coefficients the same, except the weight of good days and the weight
of bad days. We increased $a_1$ to -2,000, and $a_2$ to 2,000. We would like to see whether we could have more good
days and fewer bad days if we were willing to travel more. The best tour after this run, however, was
exactly the same as the one in the first run. Still, there was noticeably one immediate tour with 8
good days and only 2 bad days, but it had a very high number of miles (10,921). This tour still had
all zero violations.

For the third run, we set $a_1$ and $a_2$ back to their values in the first run, whereas we increased the
weight of miles to 1,000. We would like to see whether we could get fewer miles. However, the best tour
after this run was still the one in the first run, and there was no tour with fewer total mile distance
without violating any constraints.

In the next run, we increased the absolute value of the weights of good days and bad days
significantly to 20,000. Noticeably, we got one tour with 9 good days and only 2 bad days, but it had
two 1-day separation violations and a high total mile distance (9,904). The best tour after this run had
two type 1 availability violations, two type 1 separation violations, 8 good days, 2 bad days, and a
total miles of 7,960.

We also experimented different values for the parameters of simulated annealing. In the next run, we started
at a lower temperature ($t_0 = 2,500$), while keeping all other parameters, weights, and penalties the
same as those in the first run. It turned out no difference. In the next two runs, we also reduced the
number of iterations to 2,500, and then reduced ALPHA to 0.8 to decrease the temperature more
rapidly. Nevertheless, the results remained the same.

The appendix listed the manual tour and best tours that we found. Compared with the manual
tours created by the client, a lot of tours  generated by our program had desirable
characteristics:
\begin{itemize}
	\item Some of our tours had a lower number of bad days and a greater number of good days, while keeping the amount of
miles equal the amount in the manual tours (like tour 2 in the Appendix).
	\item Some of our tours increased the number of good days and decreased the number of bad days significantly, while
requiring the band to arrive at two cities only 2 days earlier than desired (like tour 6 in the
Appendix)
\end{itemize}

Moreover, the list of tours in the appendix also gave the client the ability to see  trade-offs when choosing the tours. For
example, if the client wanted to reduce one bad day from tour 2, the band would have to travel 600
additional miles. If the client wanted to maximize the number of good days while keeping the number of
bad days the same, the band would have to travel faster. This gave the client a flexibility in choosing
the tour to meet the real situation of the business.

In our opinion, the tours with all zero violations were the most preferred. Therefore, our original
choice of weights, penalties, and simulated annealing parameters was the best (tour 2 in the
Appendix).

\section{Conclusion}

The paper presents a heuristic method to create a tour for a musical band to perform in various
cities over a period, subject to many constraints and objectives. Our program first created an
initial solution and then improved this solution by simulated annealing algorithm. As it was technically
impossible to construct the initial solution that satisfied all constraints, we fulfilled as many of
them as possible and then penalized the violations in the cost function. The simulated
annealing method was simple to implement, and it produced desirable results: Given the same input
data, all produced tours had fewer violations and achieved better objectives than the manual tour
created by the client. Additionally, our program enabled the client to acknowledge trade-offs among
objectives and constraints, whose relative significance may vary over time.

Further studies may try different heuristic methods to construct a better initial solution. Although
our algorithm was designated specifically for the concert tour, the heuristic method developed here
may be reused to solve similar sequencing problems in different industries.

\newpage

\begin{appendix}

\begin{large} 
\flushleft
\bf Appendix: Manual Tour created by Client and Best Tours created by the project
\end{large}
\par
\begin{enumerate} 

\item Manual tour with zero violation, few good days, many bad days, low miles. \par
\noindent [0, 0, 0, 0, 0, 1, 0, 2, 0, 0, 3, 0, 4, 0, 5, 6, 0, 0, 0, 0, 7, 8, 0, 9, 0, 0, 0, 0, 10, 11, 0, 12, 13, 14, 15, 0, 0, 0,
0, 0, 0, 0]\par
\bigskip
\noindent Properties: \\
Good days: 4\\
Bad Days: 6\\
Number of cities in the tour: 15\\
Total miles: 7960\\
Availability violation Type 1: 0\\
Availability violation Type 2: 0\\
Break violation: 0\\
Separation violation 1 day: 0\\
Separation violation more than 1 day: 0 \par

\noindent Schedule:\\
Sat 19-Oct, Miami\\
Mon 21-Oct, Tampa\\
Thu 24-Oct, Atlanta\\
Sat 26-Oct, New Orleans\\
Mon 28-Oct, Houston\\
Tue 29-Oct, Grand Prairie\\
Sun 3-Nov, Los Angeles\\
Mon 4-Nov, San Francisco\\
Wed 6-Nov: Seattle\\
Mon 11-Nov: Chicago\\
Tue 12-Nov: Detroit\\
Wed 14-Nov: Washington DC\\
Fri 15-Nov: Philly\\
Sat 16-Nov: New York\\
Sun 17-Nov: Boston

\item Tour with zero violation, low miles (<8,000):\par
\noindent [0, 0, 0, 0, 1, 0, 0, 0, 0, 0, 2, 3, 4, 0, 5, 0, 6, 0, 0, 0, 7, 8, 0, 9, 0, 0, 0, 0, 10, 0, 11, 0, 12, 13, 14, 0, 0, 0, 0,
15, 0, 0]\par
\noindent Properties: \\
Good days: 6 \\
Bad Days: 3 \\
Number of cities in the tour: 15 \\
Total miles: 7960 \\
Availability violation Type 1: 0 \\
Availability violation Type 2: 0\\
Break violation: 0\\
Separation violation 1 day: 0\\
Separation violation more than 1 day: 0\par

\noindent Schedule:\\
Wed 16-Oct, Miami\\
Thu 17-Oct, Tampa\\
Sat 19-Oct, Atlanta\\
Sun 20-Oct, New Orleans\\
Mon 21-Oct, Houston\\
Thu 24-Oct, Chicago\\
Fri 25-Oct, Detroit\\
Sun 27-Oct, Philly\\
Mon 28-Oct, New York\\
Tue 29-Oct, Boston\\
Wed 30-Oct, Washington DC\\
Fri 1-Nov, Grand Prairie\\
Sun 3-Nov, Los Angeles\\
Mon 4-Nov, SF - Warfield\\
Wed 20-Nov, Seattle

\item Tour with zero violation, moderate miles (between 8,000 and 10,000), few bad days: \par
\noindent [0, 0, 0, 0, 1, 0, 0, 0, 0, 0, 2, 3, 4, 0, 5, 0, 6, 0, 0, 0, 7, 8, 0, 9, 0, 0, 0, 0, 0, 0, 10, 0, 12, 13, 14, 0, 0, 15, 0,
0, 11, 0]\par
\noindent Properties: \\
Good days: 6\\
Bad Days: 2\\
Number of cities in the tour: 15\\
Total miles: 8710\\
Availability violation Type 1: 0\\
Availability violation Type 2: 0\\
Break violation: 0\\
Separation violation 1 day: 0\\
Separation violation more than 1 day: 0\par
\noindent Schedule:\\
Fri 18-Oct, Miami\\
Thu 24-Oct, Tampa\\
Fri 25-Oct, Atlanta\\
Sat 26-Oct, New Orleans\\
Mon 28-Oct, Houston\\
Wed 30-Oct, Grand Prairie\\
Sun 3-Nov, Los Angeles\\
Mon 4-Nov, SF - Warfield\\
Wed 6-Nov, Seattle\\
Wed 13-Nov, Chicago\\
Fri 15-Nov, Washington DC\\
Sat 16-Nov, Philly

\item Tour with zero violation, many good days, few bad days, but high miles (>10,000): \par
\noindent [5, 0, 6, 0, 3, 0, 0, 0, 0, 1, 2, 0, 4, 0, 0, 0, 7, 0, 0, 9, 0, 8, 0, 0, 0, 0, 11, 0, 0, 0, 10, 0, 13, 14, 0, 0, 0, 0, 0,
15, 12, 0]\par
\noindent Properties: \\
Good days: 8\\
Bad Days: 2\\
Number of cities in the tour: 15\\
Total miles: 10921\\
Availability violation Type 1: 0\\
Availability violation Type 2: 0\\
Break violation: 0\\
Separation violation 1 day: 0\\
Separation violation more than 1 day: 0\par
\noindent Schedule:\\
Mon 14-Oct, Houston\\
Wed 16-Oct, Grand Prairie\\
Fri 18-Oct, Atlanta\\
Wed 23-Oct, Miami\\
Thu 24-Oct, Tampa\\
Sat 26-Oct, New Orleans\\
Wed 30-Oct, Los Angeles\\
Sat 2-Nov, Seattle\\
Mon 4-Nov, SF - Warfield\\
Sat 9-Nov, Detroit\\
Wed 13-Nov, Chicago\\
Fri 15-Nov, Philly\\
Sat 16-Nov, New York\\
Fri 22-Nov, Boston\\
Sat 23-Nov, Washington DC

\item Tour with many good days, few bad days, moderate miles, but two type-1 violations:\par
\noindent [5, 0, 0, 0, 3, 0, 0, 0, 0, 1, 2, 0, 4, 6, 0, 0, 7, 0, 0, 9, 0, 8, 0, 0, 0, 10, 11, 0, 0, 0, 0, 12, 13, 14, 0, 0, 0, 0, 0,
15, 0, 0]\par
\noindent Properties: \\
Good days: 8\\
Bad Days: 2\\
Number of cities in the tour: 15\\
Total miles: 9927\\
Availability violation Type 1: 0\\
Availability violation Type 2: 0\\
Break violation: 0\\
Separation violation 1 day: 2\\
Separation violation more than 1 day: 0\par
\noindent Schedule:\\
Mon 14-Oct, Houston\\
Fri 18-Oct, Atlanta\\
Wed 23-Oct, Miami\\
Thu 24-Oct, Tampa\\
Sat 26-Oct, New Orleans\\
Sun 27-Oct, Grand Prairie\\
Wed 30-Oct, Los Angeles\\
Sat 2-Nov, Seattle\\
Mon 4-Nov, SF - Warfield\\
Fri 8-Nov, Chicago\\
Sat 9-Nov, Detroit\\
Thu 14-Nov, Washington DC\\
Fri 15-Nov, Philly\\
Sat 16-Nov, New York\\
Fri 22-Nov, Boston\par
\noindent Separation Violation:\\
It normally takes 2.0 days to travel from New Orleans (city 4) to Grand Prairie (city 6)\\
Now it takes 1 days\\
It normally takes 5.0 days to travel from SF – Warfield (city 8) to Chicago (city 10)\\
Now it takes 4 days

\item Tour with most good days, few bad days, moderate miles, but two type-1 violations:\par
\noindent [5, 0, 2, 0, 0, 1, 0, 0, 0, 0, 0, 3, 4, 6, 0, 0, 7, 0, 0, 9, 0, 8, 0, 0, 0, 10, 11, 0, 0, 0, 0, 12, 13, 14, 0, 0, 0, 0, 0,
15, 0, 0] \par
\noindent Properties: \\
Good days: 9\\
Bad Days: 2\\
Number of cities in the tour: 15\\
Total miles: 9904\\
Availability violation Type 1: 0\\
Availability violation Type 2: 0\\
Break violation: 0\\
Separation violation 1 day: 2\\
Separation violation more than 1 day: 0\par

\noindent Schedule:\\
Mon 14-Oct, Houston\\
Wed 16-Oct, Tampa\\
Sat 19-Oct, Miami\\
Fri 25-Oct, Atlanta\\
Sat 26-Oct, New Orleans\\
Sun 27-Oct, Grand Prairie\\
Wed 30-Oct, Los Angeles\\
Sat 2-Nov, Seattle\\
Mon 4-Nov, SF - Warfield\\
Fri 8-Nov, Chicago\\
Sat 9-Nov, Detroit\\
Thu 14-Nov, Washington DC\\
Fri 15-Nov, Philly\\
Sat 16-Nov, New York\\
Fri 22-Nov, Boston\par
\noindent Separation Violation:\\
It normally takes 2.0 days to travel from New Orleans (city 4) to Grand Prairie \hbox{(city 6)}\\
Now it takes 1 days\\
It normally takes 5.0 days to travel from SF – Warfield (city 8) to Chicago (city 10)\\
Now it takes 4 days\\

\end{enumerate}

\end{appendix}

\end{document}